\documentclass[sigconf]{acmart}

\usepackage{framed}
\usepackage{multirow}
\usepackage{listings}

\AtBeginDocument{%
  }

\copyrightyear{2025}
\acmYear{2025}
\setcopyright{rightsretained}
\acmConference[CHI EA '25]{Extended Abstracts of the CHI Conference on Human Factors in Computing Systems}{April 26-May 1, 2025}{Yokohama, Japan}
\acmBooktitle{Extended Abstracts of the CHI Conference on Human Factors in Computing Systems (CHI EA '25), April 26-May 1, 2025, Yokohama, Japan}\acmDOI{10.1145/3706599.3720189}
\acmISBN{979-8-4007-1395-8/2025/04}

\begin{document}

\title[\sysname{}: Broadening Design Exploration with Text-to-Image Model]{\sysname{}: Broadening Design Exploration with\\Text-to-Image Model}

\author{DaEun Choi}
\email{daeun.choi@kaist.ac.kr}
\affiliation{%
  \institution{KAIST}
  \city{Daejeon}
  \country{Republic of Korea}
}

\author{Kihoon Son}
\email{kihoon.son@kaist.ac.kr}
\affiliation{%
  \institution{KAIST}
  \city{Daejeon}
  \country{Republic of Korea}
}

\author{Hyunjoon Jung}
\email{hjung@adobe.com}
\affiliation{%
  \institution{Adobe}
  \city{San Jose, CA}
  \country{United States}
}

\author{Juho Kim}
\email{juhokim@kaist.ac.kr}
\affiliation{%
  \institution{KAIST}
  \city{Daejeon}
  \country{Republic of Korea}
}

\newcommand{\sysname}[0]{Expandora}

\begin{abstract}
  Broad exploration of references is critical in the visual design process. While text-to-image (T2I) models offer efficiency and customization of exploration, they often limit support for divergence in exploration. We conducted a formative study (N=6) to investigate the limitations of current interaction with the T2I model for broad exploration and found that designers struggle to articulate exploratory intentions and manage iterative, non-linear workflows. To address these challenges, we developed \sysname{}. Users can specify their exploratory intentions and desired diversity levels through structured input, and using an LLM-based pipeline, \sysname{} generates tailored prompt variations. The results are displayed in a mindmap-like interface that encourages non-linear workflows. A user study (N=8) demonstrated that \sysname{} significantly increases prompt diversity, the number of prompts users tried within a given time, and user satisfaction compared to the baseline. Nonetheless, its limitations in supporting convergent thinking suggest opportunities for holistically improving creative processes.


\end{abstract}



\begin{teaserfigure}
  \includegraphics[width=\textwidth]{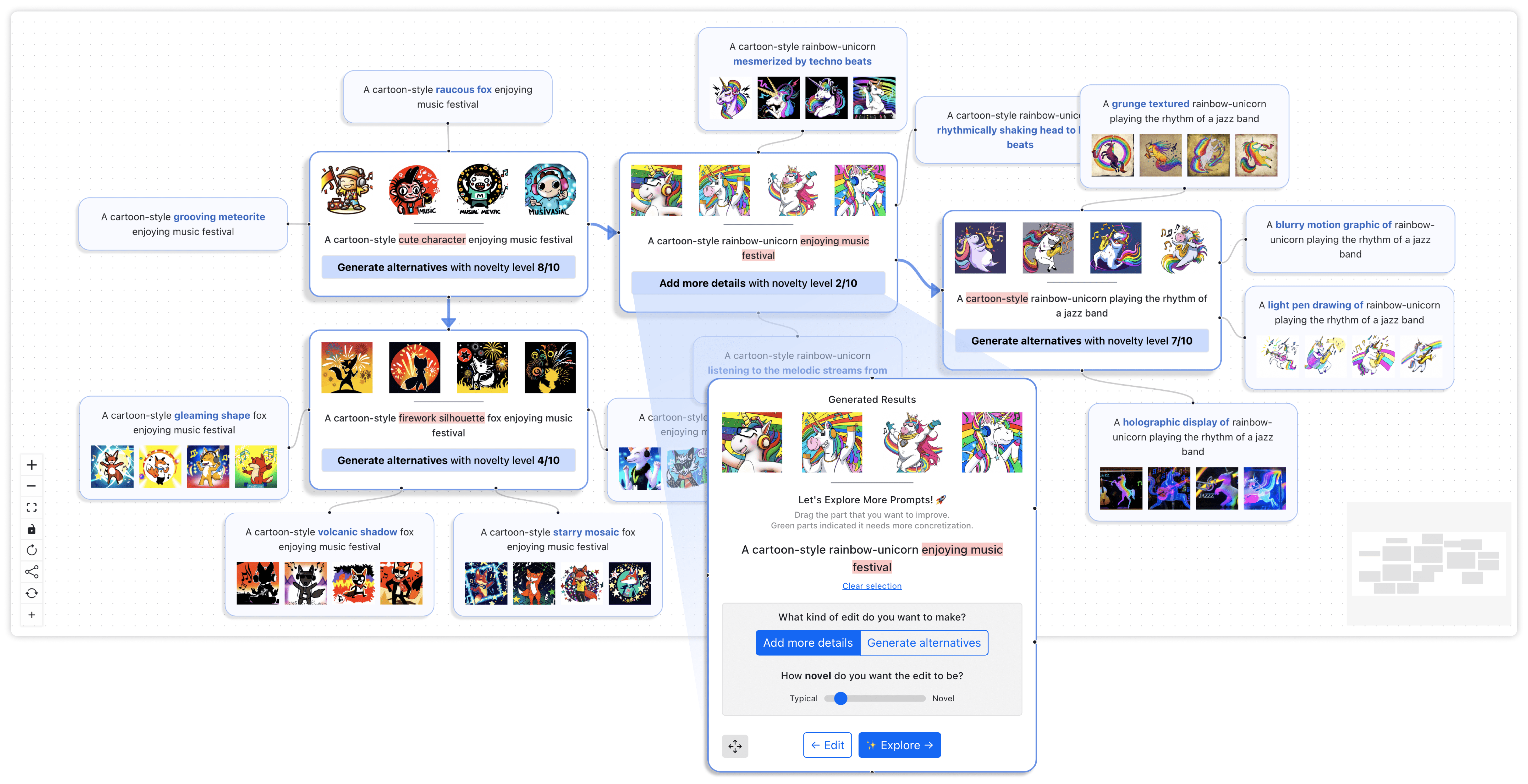}
  \caption{Interface of \sysname{}. There are two key components of \sysname{}: (1) Structured Input Interface: Users can specify exploratory intentions by selecting parts of a prompt to refine, choosing between adding details or generating alternatives, and adjusting the desired novelty level. (2) Mindmap-like Interface Showing Results: Exploration results are visualized as a branching, non-linear structure that reflects iterative workflows, allowing users to revisit, refine, or branch off ideas.}
  \label{fig:teaser}
\end{teaserfigure}


\maketitle

\section{Introduction}
Exploration is an essential first step in the creative process~\cite{eckert2000sources}. By exploring a wide range of references, people understand the problem space and the landscape of existing ideas and finally generate novel ideas~\cite{herring2009getting, lee2010designing, ritchie2011d, muller2011leaving}.
For graphic designers, traditional tools like reference websites (e.g., Pinterest) have long supported this exploratory process, but text-to-image (T2I) models have emerged as alternative sources of references. These models enable designers to generate images quickly, tailored to their specific needs. Also, designers often expect that models can surprise them with creative images beyond their initial expectations~\cite{ko2023creativeworks}.
However, when we look at how users perceive outputs and develop ideas, they can limit the \textit{divergence} of exploration.
The model generates a predictable range of images or leads to fixation on initial outputs~\cite{ko2023creativeworks, wadinambiarachchi2024generative}, limiting the breadth of exploration.

To better understand the underlying causes of these limitations, we conducted a formative study with 6 participants to investigate the challenges of using T2I models during exploration, focusing on the evolution of their ideas and intentions over time. Our findings revealed two key challenges. First, participants struggle to express their \textit{exploratory intentions} effectively through text prompts, particularly specifying which design elements to vary, which to keep consistent, and the desired level of diversity. Second, the sequential interface of current T2I systems fails to support the non-linear and iterative nature of the exploration activities, such as branching into multiple ideas or revisiting previous iterations, which further hinders effective ideation.

Based on our findings, we developed \sysname{} for supporting T2I-assisted design exploration by enabling users to explore a wide variety of semantically diverse and novel prompts. To help users articulate their exploratory intentions more directly, \sysname{} provides a structured input interface where users can specify which aspects to explore, how to explore, and what level of diversity they want. Then, using an LLM-based pipeline, the system generates a range of prompt variations tailored to the user's intention. The results are presented in a mindmap-like interface, which can organize ideas hierarchically and allow users to explore branching and iterative pathways visually.


To evaluate \sysname{}, we conducted a within-subjects comparative study with 8 participants familiar with graphic design and the T2I model. We compared \sysname{} to ChatGPT with an image generation feature. Results showed that \sysname{} enabled participants to try semantically more diverse prompts into T2I models, showing that \sysname{} facilitated more diverse exploration. Also, participants explored a significantly greater number of prompts with \sysname{} within the same amount of time compared to the baseline and reported higher satisfaction with their exploration outcomes.
While \sysname{} effectively supported divergent exploration, participants noted that creative design requires both divergent and convergent processes. They expressed a need for features supporting convergent ideation to refine explored ideas into detailed outputs. We discuss potential improvements to \sysname{} to address this gap and better support the overall creative process.
\section{Background}

\subsection{Creativity Support in the Idea Exploration Process}

In early-stage idea exploration, users must balance divergent and convergent thinking to reach more successful creative outcomes~\cite{frich2021how, chong2025prompting}. To support the creative exploration process, prior work explored seamless and semantically connected interfaces that allow users to create links to visual and semantic data in the ideation process~\cite{koch2020semanticcollage, kang2021metamap}. With the advancement of generative models (e.g., LLMs or diffusion models), there is another research stream that explores how to use these models to support the creative ideation process. CreativeConnect~\cite{choi2024creativeconnect} and PopBlends~\cite{wang2023popblends} support graphic designers in the idea decomposition and recombination process. Also, Luminate~\cite{suh2024luminate} supports users' divergent and convergent thinking when interacting with LLMs to avoid fixation. Building on these two main streams, we propose \sysname{} that integrates a semantic node-based interface with generative models to support both divergent and convergent thinking in design exploration.

\subsection{Intent Expression with Generative Model}
Supporting accurate intent expression is crucial in the exploration stage since the user needs to express two types of intent: specification (e.g., refining a clear idea or output in mind) and diversification (e.g., exploring broad or unclear possibilities with a certain direction). Recent work has shown how to utilize the generative model's ability as an expression channel of a user's intent ~\cite{son2024genquery}. For input methods, various generative model-based tools~\cite{wang2024promptcharm, peng2024designprompt, zhou2024stylefactory} support structured or tailored inputs beyond natural language input to help users express their creative intention. Beyond providing various input modalities, recent work also suggested the method of automatically refining and suggesting users' prompts to enhance their exploration~\cite{brade2023promptify, son2024genquery, almeda2024prompting}. In alignment with these research directions, our work includes 1) a structured intent input approach with 2) guidance on how to diversify \& specify prompts.
\section{Formative Study}

We conducted a formative study to investigate the challenges designers face while exploring and evolving ideas using T2I models.
We recruited six design students or graduates with experience in at least three graphic design projects. Participants were asked to brainstorm about the mascot character design for their company (for practitioners) or department (for students) using image generation models for a 40-minute session. Each time they prompted the model to generate an image, they were asked to explain their current intentions and expectations for the outcome. After reviewing the generated images, they reflected on how well the results aligned with their intentions.
Following the session, we conducted a 15-minute semi-structured interview to delve deeper into the challenges participants encountered throughout the process and to explore potential improvements to the user experience.

\subsection{Findings}
Participants reported two major limitations of interacting with the T2I model during the design exploration.

First, they \textbf{struggled to express \textit{exploratory} intentions accurately through text prompts}. Design exploration often begins with an open-ended mindset, where users have broad ideas and seek to explore diverse possibilities through generated images. However, participants found it difficult to communicate this intention about desired diversity to the model. Based on the observations, we define the user's intention during this phase as \textit{exploratory intentions}, focusing on \textit{how much and what kind of diversity} they want in the outputs. 
We also found that this intention of diversity often varies across image elements—some may need to remain consistent, while others can vary widely. In this case, articulating this intention into a text prompt becomes much harder. For example, one participant shared: \textit{``When I typed 'a drawing of a soft cloud,' I wanted the shape of the cloud to vary, but I wanted the drawing style to remain consistent with the previous one. However, I couldn’t specify these preferences to the model.''}
Current interfaces for T2I models lack inputs to deliver such exploratory intentions when the users don't have a clear goal. As a result, users are often forced to either specify their prompts before fully exploring diverse ideas or use vague prompts and rely on the model’s randomness, which may not align with their creative goals.

Second, they \textbf{struggled to engage in iterative design ideation using linear interfaces in current T2I systems}.
Participants highlighted the discrepancy between the nature of the design ideation process and the current T2I interface. The design ideation inherently involves non-linear actions, such as branching into multiple ideas, merging concepts, or reverting to previous iterations. However, the interface of the current T2I systems is linear, such as a sequential generation history, which makes it difficult to revisit or combine ideas during the process.
P3 described this as: \textit{"I usually repeat exploration and then select parts to refine. But with generative models, it felt like I was going back to square one with every generation."} This linearity constrains the iterative nature of creative workflows, making it challenging for users to manage their ideas or build upon prior explorations.

Based on the findings, we derived two design goals for the system to address the challenges faced by designers during the use of the T2I models for design exploration:

\begin{itemize}
    \item {DG1.} \textbf{Support Input for Exploratory Intention}: Enable users to specify which elements to explore, and provide controls to define the direction and range of the diversity of the T2I outputs. 
    This ensures that users can explore broad possibilities aligned with their exploratory intentions.
    \item {DG2.} \textbf{Facilitate Iterative and Non-Linear Workflows}: Support iterative exploration by enabling users to explore on top of previous exploration, revisit exploration history, or branch ideas seamlessly, fostering a more natural and flexible creative process.
\end{itemize}
\section{\sysname{}}

Based on the design goals, we developed \sysname{}, a web-based system that supports users to explore diverse and novel design ideas using a T2I model with an LLM-based pipeline.

\subsection{Features}

\begin{figure*}
    \centering
    \includegraphics[width=1\linewidth]{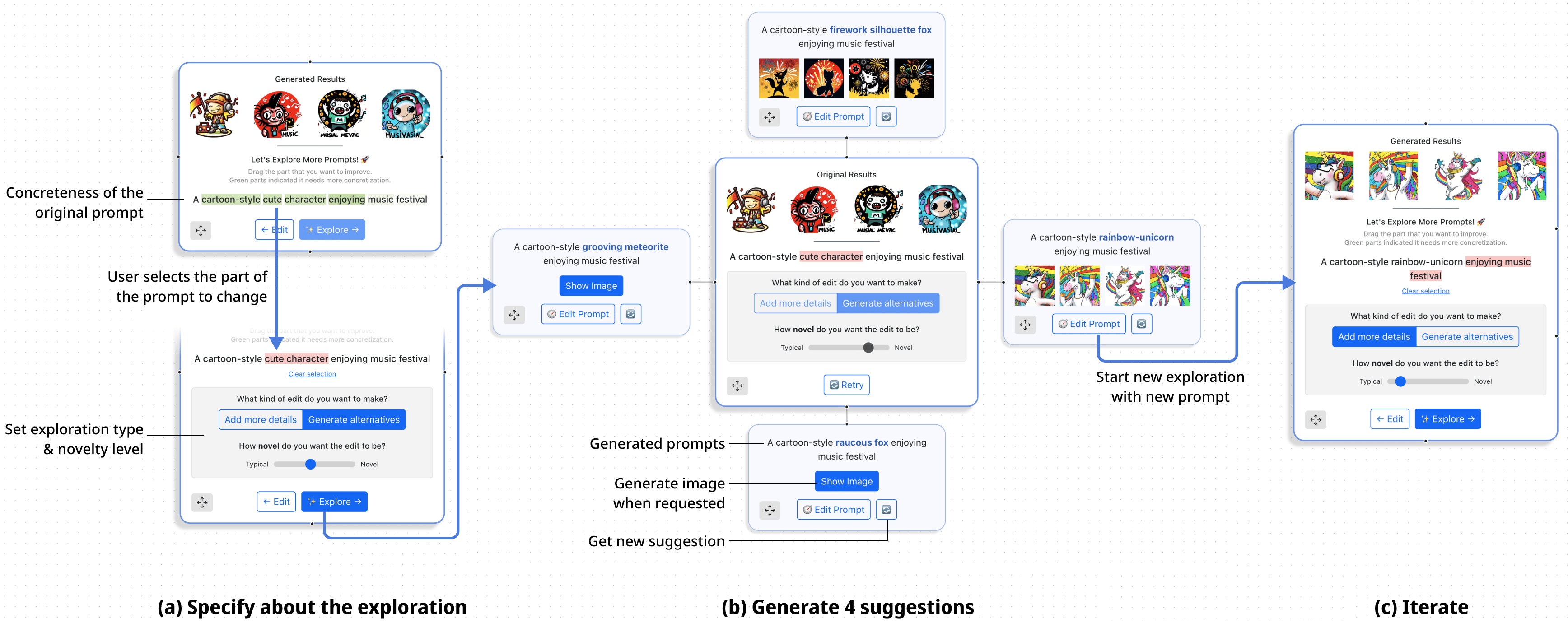}
    \caption{Usage flow of \sysname{}. (a) The user begins by entering a prompt, and the system generates four initial images. Additionally, the system highlights the concreteness of each word in the prompt. The system also provides tools for further diversification, allowing users to select a specific part of the prompt to refine, choose how to diversify it, and adjust the desired level of novelty. (b) Based on the input, the system suggests four modified prompts that refine the specified part. Users can either select a prompt to generate new images or request additional suggestions. (c) For further exploration, users can click "Edit Prompt" to return to the interface in (a) and continue diversifying, enabling iterative and targeted exploration.}
    \label{fig:system-flow}
\end{figure*}

\subsubsection{Structured Input for Exploration Intention}
As shown in Figure \ref{fig:system-flow} (a), the system workflow begins with users entering an initial prompt, and the system generates four images based on it. To assist users in choosing what to explore, the system evaluates the concreteness of each word in the prompt.
Previous research~\cite{leake2020generating} uses concreteness as a measure of how much the words are connected to specific visual representations. Drawing on this, we use concreteness as an indicator of whether each part of the prompt is specific enough or requires further elaboration. Words are highlighted in green, with thicker highlights indicating less concrete terms that need refinement. How to calculate the concreteness is described in Section~\ref{sec:tech}.

The system provides users with structured input for customizing where and how to explore further, designed based on the observation of user intentions from the formative study. First, by referring to the concreteness, users can specify the part of the prompt they wish to explore more by dragging on it. Then, they can choose to either add details or generate alternative variations. They can also set the level of novelty, with lower levels producing focused outputs that are highly connected to the original prompt, and higher levels yielding diverse and loosely connected ideas. These features allow users to articulate their exploration direction and breadth (DG1).

\subsubsection{System Suggestions for the Expanded Prompts}

Based on the user's settings, the system generates four prompt suggestions that modify the selected part of the prompt (Figure \ref{fig:system-flow} (b)). These suggestions are displayed in a mindmap-like layout. Each suggestion initially shows a modified version of the prompt. When the user clicks the "Show Image" button, the system generates four images based on the modified prompt. If a suggestion is unsatisfactory, users can click a "New Suggestion" button, and the system recalculates suggestions using an adaptive algorithm (detailed in Section~\ref{sec:tech}). This ensures that subsequent suggestions better align with the user’s preferences.

\subsubsection{Iterative Exploration Workflow}

When users click "Edit Prompt" on each suggestion, they can repeat the exploration process on top of the suggested prompt: the system allows users to set exploration preferences and generates four new suggestions. This iterative approach enables users to refine ideas step by step, with each iteration expanding the design space. The exploration history is shown in a mindmap-like interface (shown in Figure~\ref{fig:teaser}). This supports non-linear workflows by allowing users to branch into multiple ideas or revisit prior prompts at any point in their exploration journey. By visually representing these connections, the system helps users manage complex ideation processes and facilitates a flexible workflow that mirrors natural design ideation practices (DG2).

\subsection{Technical Detail} \label{sec:tech}

\sysname{} was built as a web-based application using ReactJS\footnote{https://react.dev/} and React Flow\footnote{https://reactflow.dev/}. We used OpenAI's DALL·E 2\footnote{https://openai.com/index/dall-e-2/} API for image generation. We developed ML pipelines to calculate the concreteness of prompts and generate prompt suggestions based on the user's structured input, with technical details discussed in subsequent sections.

\subsubsection{Concreteness}

To assess the concreteness of each word in image generation prompts, we trained a BERT~\cite{devlin2019bertpretrainingdeepbidirectional}-based concreteness model. We used a 40k-word concreteness rating dataset by Brysbaert et al.~\cite{brysbaert2014concreteness} and achieved a loss of 0.042 and a Mean Absolute Error (MAE) of 0.346. The results were visualized with green highlights: lower scores appeared as more opaque green, while scores closer to the maximum of 5 became increasingly transparent.

\subsubsection{Prompt Suggestion Algorithm}
\textbf{Generating the suggestions}.
The system uses GPT-4o to generate 200 prompt suggestions. We designed two prompts to support the two different ways of prompt diversification (adding detail / generating alternatives). For ``add more details'', the LLM is instructed to generate diverse ways to add richer descriptions on the selected part (Appendix-\ref{appendix:detailprompt}). For ``generate alternatives'', it provides diverse words or phrases that can replace the user-specified part, considering the other parts of the prompt (Appendix-\ref{appendix:alternativeprompt}).

\textbf{Filtering the first suggestions}.
After generating 200 prompts, the system filters them based on semantic similarity to align with the user's desired level of novelty (0-1). Prompts with similarity scores falling within the range \(\text{novelty\_input} \pm 0.2\) relative to the original prompt are retained.
To provide four representative suggestions, the filtered prompts are clustered based on their semantic similarity to each other using the K-means algorithm. The centroids of the four clusters, representing the most central and distinctive examples, are selected as the final suggestions for the user.

\textbf{Alternative suggestions following the removal input.}
When the user removes a suggestion, it indicates dissatisfaction with that option. Therefore, the system selects a new cluster centroid that is both maximally distant from the removed suggestion and sufficiently distinct from the existing centroids. The new centroid is determined by maximizing its combined distance from the removed text and the current centroids.
\section{Evaluation}

We conducted a within-subjects comparative study (N=8) to assess the effectiveness of \sysname{} in supporting design exploration using the T2I model. We provided ChatGPT with an image generation feature as a baseline, as it is one of the most powerful and widely used tools for ideating and generating images. All 8 participants (6 females, 2 males; age M=23.75 and SD=1.67) reported having prior experiences participating in three different design projects and regularly using image generation models for their design process. Details about the participants are in Appendix~\ref{appendix:participant}.

Participants completed the design exploration task twice for brainstorming various concepts for (1) \textit{A mascot character for a multi-genre music festival with an exotic vibe} and (2) \textit{A mascot character for a laboratory researching cutting-edge future technology}. Each task lasted 25 minutes, preceded by a 5-minute tutorial on the assigned tool (\sysname{} or ChatGPT). The order of topics and systems used was counterbalanced across participants. After each session, they completed the post-task survey. After both sessions, we conducted 15-minute semi-structured interviews.

\section{Results}

Results showed that \sysname{}'s structured input channel for users to express their exploratory intention was helpful for efficient exploration. \sysname{}'s prompt suggestions and the mindmap-like interface were also helpful for diversification. Full survey results are shown in Appendix~\ref{appendix:fullsurvey}. We used a Wilcoxon signed-rank test for survey questions, as they were ordinal data on a 7-point Likert scale. For the usage log analysis, we conducted a two-sample paired t-test.

\subsection{Effective Exploration, Higher Satisfaction} \label{sec:efficiency}


During the 25-minute design exploration task, participants using \sysname{} tested significantly more prompts (M=30.88, SD=9.16) compared to the baseline (M=14, SD=4.72; p=0.0007). Despite this increased activity, NASA-TLX~\cite{HART1988139} results showed no significant difference in overall workload between the two systems. Furthermore, participants reported lower temporal demand with \sysname{}, indicating that its features streamlined their workflow and reduced the perceived time pressure.
As a result, participants reported significantly higher satisfaction with their exploration results when using \sysname{} (M=6.13, SD=0.64) compared to the baseline (M=5.25, SD=0.89; p=0.038).

Interview results revealed that this advantage was mainly driven by \sysname{}'s structured input. Participants highlighted that it enabled them to express their intentions easily when their ideas are unspecified, without requiring additional effort to refine or concretize them. As P3 explained, \textit{``Even when I had no idea what to create, the system gave me something to try.''} Participants also mentioned that the system helped them develop a clear mental model of how diversification will work, which gave them a sense of control over the process. This efficiency in interaction ultimately allowed participants to focus more on the act of exploration itself.


\subsection{More Diversified, but No Significant Creativity Gains}


Participants said that the prompts generated by \sysname{} were novel. For the survey asking whether the ideas explored through the system were predictable and obvious, participants disagreed more when using \sysname{} (M=1.75, SD=0.71) than those from the baseline (M=3.88, SD=1.36; p=0.017).
This led users to explore semantically more diverse ideas in the overall process. By analyzing what prompts users used to generate images, it was shown that prompts in the \sysname{} condition were more diverse with lower average semantic similarity between prompts (M=0.476, SD=0.10) compared to the baseline (M=0.666, SD=0.10; p=0.0032).
As a result, the generated images in \sysname{} condition were also diverse, as shown in Appendix~\ref{appendix:example}.

Despite this advantage in terms of broadening the exploration space, there were no differences in the Creativity Support Index (CSI)~\cite{csi} between the two conditions, indicating that diversification alone was insufficient to support end-to-end creative exploration compared to the baseline.
Interview findings also aligned with this result. While 7 out of 8 participants preferred \sysname{} over the baseline for \textit{diverse} exploration, only 5 preferred it for \textit{overall creative} exploration. Participants explained that creativity requires both divergent and convergent processes, but \sysname{} primarily supports divergence, leaving the convergent aspect of creativity largely unsupported.

\subsection{Impact on User’s Exploration Process}

Participants' exploration patterns differed distinctly between \sysname{} and baseline. This section describes two core differences that we identified through the interview.

\subsubsection{Divergence-centric Exploration}

With baseline (ChatGPT), participants typically began by asking for a list of ideas, selecting 2-3 of them, and then sequentially refining them. Therefore, most of the processes were convergent. In contrast, \sysname{} fostered a more divergence-centric approach, where participants started with a single point but continued to shift between variations and explore new directions without reaching a definitive conclusion.

Ultimately, this process encouraged participants to think more openly about their goals. Participants noted that they kept trying different things without a single strong intention compared to the baseline.
However, this divergent focus often made participants fail to converge into final ideas.

\subsubsection{More Reflection and Control}

Participants noted that \sysname{}'s non-linear interface encouraged deeper reflection of their exploration process and gave them better control over it. Participants said that the interface facilitated comparing different suggestions, letting participants think about their preferences and goals. P6 explained, \textit{``By selecting between 4 suggestions and refining them, I was able to reflect on what I truly wanted and adjust my direction accordingly.''} Moreover, by showing the whole journey of the idea's evolution, participants said that they could easily identify what actions they have to take next, such as reverting to a specific point in their design process or determining which direction to explore. As P5 explained, \textit{``Seeing my ideation process visually helped me identify points that I need to revisit or refine to accomplish my goal.''} 
\section{Discussion} \label{sec:discussion}

We examined how \sysname{}'s design supports users to explore diverse design through structured input and non-linear interfaces. This section discusses its impact on the creative process including both divergence and convergence, and potential improvements for future systems.

\subsection{Structured Input for Divergence and Convergence in Exploration}

In divergent thinking, quickly exploring a large amount of different options is crucial. To support this, the structured input of \sysname{} was designed to enhance efficiency in trying out diverse exploration. Based on the formative studies, we identified two core actions---adding detail and generating alternatives---paired with a novelty slider. These features effectively improved efficiency during divergence, as shown in Section~\ref{sec:efficiency}. This approach could be also adapted to other creative domains with different modalities, such as storytelling (text) or songwriting (audio), by analyzing the actions and qualities users prioritize during exploration.

However, creative exploration also involves convergence, where refining and detailing are required. Prior research~\cite{chong2025prompting} emphasizes that successful design ideation requires a transition from broad exploration to structured refinement. \sysname{} lacked channels to deliver convergent intent, causing frustration and difficulties in steering outputs as desired in this convergent phase, reflected in lower ratings for \textit{accurately expressing exploration direction} in the survey (\sysname{}: M=4.13, SD=1.46 / Baseline: M=5.86, SD=1.13; p=0.006).
Future systems should incorporate convergent support in addition to \sysname{}'s current design. One way could be incorporating free-form inputs alongside structured options. For instance, letting users add keywords to include or exclude in new suggestions would help them refine their focus and express creative intent more effectively. Another enhancement could involve refining outputs based on user-preferred suggestions, such as merging multiple ideas, to better support users to converge faster. Additionally, tools could provide these functionalities dynamically based on the user's current mode of divergence or convergence to effectively guide each phase.



\subsection{Applying Non-Linear Interfaces to T2I Systems}

Our study showed that the non-linear interface of \sysname{} aligns better with designers' ideation process and helps reflect on and expand their exploration paths. This resulted in diverse exploration and fostered a sense of control and coherence in the process. Such non-linear interfaces can be adapted to generative AI systems in various ways. A simple enhancement could be changing the way of displaying generation history in a graph or tree-like structure, which could provide users with a clearer sense of their exploration trajectory. These visual cues would enable users to revisit previous iterations, compare results, and build upon promising directions, making the creative process more intuitive and user-friendly.

Moreover, the non-linear interface offers data about how users modify their prompt over time or what kind of suggestions users reject or develop, which can provide valuable insight into how each user usually develops and evolves their ideas. By extracting patterns of this ideation sequence, the system could understand the user's own style of exploration, which could be used to offer more targeted suggestions during subsequent exploration. Furthermore, the system could even simulate exploration on behalf of the user. These enhancements could further streamline the creative process, enabling systems to provide smarter, more context-aware support.



\section{Conclusion}

This paper introduced \sysname{}, a system designed to enhance expressive and iterative exploration using text-to-image (T2I) models. By reducing cognitive load for expressing exploratory intention through structured input interaction, \sysname{} enabled users to focus on exploring diverse ideas. The mindmap-like interface further supported reflection, comparison, and flexible navigation of their journey. The user study (N=8) showed that users with \sysname{} tried more prompts for the image generation model and explored semantically more diverse concepts than the baseline. However, the lack of convergent support highlighted areas for improvement, such as adding flexible inputs.

\begin{acks}
This work was supported by Institute of Information \& Communications Technology Planning \& Evaluation (IITP) grant funded by the Korea government (MSIT) (No.2021-0-01347,Video Interaction Technologies Using Object-Oriented Video Modeling).
This work was also supported by the National Research Foundation of Korea (NRF) grant funded by the Korea government (MSIT) (No.RS-2024-00406715)
\end{acks}

\bibliographystyle{ACM-Reference-Format}
\bibliography{main}


\appendix

\section{Technical Details}

\subsection{Prompt: Adding details}
\label{appendix:detailprompt}
\sloppy
\begin{framed}
    \footnotesize
    \texttt{
    Improve user-provided prompts for image generation by adding details to enhance specificity, clarity, and creativity. Users will submit an original prompt and indicate areas for refinement. Generate revisions for the part, mainly by adding more details for the specified part.\\\# Steps\\1. First, analyze the user's original prompt to understand the overall theme and concept mentioned.\\2. Review the part specified by the user that needs improvement. Take note of ambiguities, lack of detail, or potential enhancements in this part.\\3. Generate 200 variations.\\- 100 Literal Revisions: Standard, precise, and expected descriptions.\\- 100 Creative Revisions: Unique, imaginative, and highly inventive descriptions.\\\# Input Format\\1. Original Prompt: [Insert the original prompt here].\\2. Part to Change: [Specify the part of the prompt to modify].\\3. Index of the Part: [Specify the start and end index of the part to modify].\\\# Output Format\\Provide a list of 200 variations without numbering.\\\# Example 1\\Original Prompt: A scientist character doing an experiment\\Part to Change: scientist character\\Index of the Part: 2-20\\Literal Revisions: scientist in a white lab coat, scientist holding a test tube, chemist adjusting a Bunsen burner, roboticist surrounded by futuristic machines, biologist analyzing DNA samples.\\Creative Revisions: scientist fused with advanced AI, scientist glowing with ethereal equations, cosmic alchemist crafting stardust, enchanted scientist wielding magical flasks, intergalactic inventor with robotic arms.\\\# Example 2\\Original Prompt: A scientist character doing an experiment\\Part to Change: doing an experiment\\Index of the Part: 22-40\\Literal Revisions: mixing chemicals in a beaker, calibrating a high-tech microscope, analyzing data on a holographic screen, testing a prototype in a lab, extracting DNA from a sample.\\Creative Revisions: manipulating glowing plasma orbs, distilling elixirs in an enchanted lab, running tests on alien life forms, experimenting with anti-gravity fields, crafting potions of futuristic energy.\\\# Notes\\- Output only consists of the modified part of the prompt.\\- Ensure revisions do not overlap or repeat details from the unspecified part of the prompt. For example, when the original prompt is \"A character walking\", and the specified part is \"character\", the revision should not include any elements that are about the character's action (which will be overlapped with \"walking\").\\- The revision should be clear and specific enough to be used for image generation prompts.\\- The revision should align well with the other part of the prompt.\\- Avoid using complex words and phrases.
    }
\end{framed}

\subsection{Prompt: Generating alternatives}
\label{appendix:alternativeprompt}
\sloppy
\begin{framed}
    \footnotesize
    \texttt{
    Generate 200 diverse and creative phrases to replace a specific part of an image generation prompt. Each replacement should offer a different entity from the original but maintain a related vibe or essence. For instance, if the original term is \"kid,\" alternatives like \"angel\" or \"puppy\" should evoke a similar feeling. Aim for variety, ensuring users find inspiration, with each replacement clear and suitable for immediate use in image prompts.\\\# Steps\\1. First, analyze the user's original prompt to understand the overall theme and concept mentioned.\\2. Examine the specific part of the prompt provided and explore related concepts that share a similar ambiance, emotion, or function as the original but have a different entity.\\3. Generate 200 variations.\\- 100 Literal Variations: Standard and easily expected variations.\\- 100 Creative Variations: Unique and highly inventive variations.\\\# Input Format\\1. Original Prompt: [Insert the original prompt here].\\2. Part to Change: [Specify the part of the prompt to modify].\\3. Index of the Part: [Specify the start and end index of the part to modify].\\\# Output Format\\Provide a list of 200 variations without numbering.\\\# Example 1\\Original Prompt: A scientist character doing an experiment\\Part to Change: scientist\\Index of the Part: 2-10\\Literal Revisions: engineer, mathematician, cute astronauts, AI developer\\Creative Revisions: deep sea explorer, time traveler, mad inventor, lunar artist. AI robot, VR monster\\\# Example 2\\Original Prompt: A scientist character doing an experiment\\Part to Change: doing an experiment\\Index of the Part: 22-40\\Literal Revisions: writing notes, asking a question, coding algorithms, recording videos, sleeping in front of a monitor, studying with a thick book\\Creative Revisions: dancing with a robot, painting an artistic picture, playing computer games, observing starts, dreaming about the future\\\# Notes\\- Ensure that each alternative maintains a balance between being distinct yet related in vibe to the original term.\\- Output only consists of the modified part of the prompt.\\- Ensure generated alternatives do not overlap or repeat details from the unspecified part of the prompt. For example, when the original prompt is \"A character walking\", and the specified part is \"character\", the alternatives should not include any elements that are about the character's action (which will be overlapped with \"walking\").\\- The revision should be clear and specific enough to be used for image generation prompts.\\- The revision should align well with the other part of the prompt.\\- Avoid using complex words and phrases.
    }
\end{framed}

\section{Details of Study Participants} \label{appendix:participant}

Details of study participants are shown in Table~\ref{tab:participant}. Participants responded to two 7-point Likert scale questions asking their familiarity with writing prompts ("I know how to write prompts to achieve the desired image generation results.") and their understanding of how T2I models function ("I have a sufficient understanding of how image generation models work.") in the enrollment survey.

\begin{table*}[t]
\begin{tabular}{cccccc}
\toprule
\textbf{PID} & \textbf{Age} & \textbf{Sex} & \textbf{Frequency of Using T2I Model} & \textbf{\begin{tabular}[c]{@{}c@{}}Prompt authoring\\ familiarity (7-scale)\end{tabular}} & \textbf{\begin{tabular}[c]{@{}c@{}}Understanding of\\ T2I model (7-scale)\end{tabular}} \\ \midrule
P1           & 26           & Female       & 1-2 times a month                 & 5                                                                                         & 6                                                                                       \\
P2           & 25           & Female       & Once a week                           & 3                                                                                         & 3                                                                                       \\
P3           & 22           & Female       & Once in 2-3 months               & 3                                                                                         & 6                                                                                       \\
P4           & 21           & Female       & 2-5 times a week                 & 7                                                                                         & 7                                                                                       \\
P5           & 23           & Female       & Almost every day                      & 5                                                                                         & 3                                                                                       \\
P6           & 24           & Female       & 1-2 times a month                 & 5                                                                                         & 2                                                                                       \\
P7           & 25           & Male         & Once a week                           & 5                                                                                         & 4                                                                                       \\
P8           & 24           & Female       & 1-2 times a month                 & 5                                                                                         & 5                                                                                       \\ \bottomrule
\end{tabular}
\caption{Details of study participants, including their age, sex, frequency of using text-to-image (T2I) models in the past six months, and survey results on prompt authoring familiarity and understanding of the T2I model}
\label{tab:participant}
\end{table*}

\section{Full Survey Results} \label{appendix:fullsurvey}

The following survey questions, provided to participants, were measured on a 7-point Likert scale:

\begin{itemize}
    \item Accurately expressing exploration direction: I was able to accurately express the direction of exploration I wanted to the system.
    \item Defining exploration scope clearly: I was able to clearly define the scope of exploration I wanted to the system.
    \item System understood intent well: I believe the system understood my exploration intent well and acted accordingly.
    \item Easily adjusted system behavior: I found it easy to adjust the system to operate in line with my exploration intent.
    \item Satisfied with overall exploration outcome: Overall, I am satisfied with the results of my exploration.
    \item Results helped creative exploration: The results generated by the system were helpful for my creative exploration.
    \item System leads to fixation: This system limits exploration to specific ideas.
    \item Ideas are obvious and simple: The ideas explored through this system are obvious and easy to think of.
\end{itemize}

\begin{table*}[t]
\begin{tabular}{ccrrrrr}
\hline
                                    &                           & \multicolumn{2}{c}{baseline}                    & \multicolumn{2}{c}{\sysname{}}                     & \multicolumn{1}{c}{\multirow{2}{*}{P}} \\ \cline{3-6}
                                    &                           & \multicolumn{1}{c}{M} & \multicolumn{1}{c}{STD} & \multicolumn{1}{c}{M} & \multicolumn{1}{c}{STD} & \multicolumn{1}{c}{}                   \\ \hline
\multicolumn{2}{c}{Accurately expressing exploration direction} & 5.88                  & 1.13                    & 4.13                  & 1.46                    & \textbf{0.016}                         \\
\multicolumn{2}{c}{Defining exploration scope clearly}          & 5.38                  & 1.06                    & 4.50                  & 1.60                    & 0.084                                  \\
\multicolumn{2}{c}{System understood intent well}               & 5.13                  & 1.36                    & 4.75                  & 0.89                    & 0.517                                  \\
\multicolumn{2}{c}{Easily adjusted system behavior}             & 3.63                  & 1.06                    & 5.13                  & 1.25                    & 0.088                                  \\
\multicolumn{2}{c}{Satisfied with overall exploration outcome}  & 5.25                  & 0.89                    & 6.13                  & 0.64                    & \textbf{0.038}                         \\
\multicolumn{2}{c}{Results helped creative exploration}         & 4.88                  & 1.64                    & 5.50                  & 1.77                    & 0.547                                  \\
\multicolumn{2}{c}{System leads to fixation}                    & 3.63                  & 1.77                    & 2.88                  & 1.55                    & 0.547                                  \\
\multicolumn{2}{c}{Ideas are obvious and simple}                & 3.88                  & 1.36                    & 1.75                  & 0.71                    & \textbf{0.017}                         \\ \hline
\multicolumn{2}{c}{Aggregated CSI}                              & 5.06                  & 0.81                    & 5.53                  & 0.80                    & 0.383                                  \\ \hline
\multirow{6}{*}{NASA-TLX}           & Mental demand             & 2.75                  & 1.75                    & 3.00                  & 1.41                    & 0.914                                  \\
                                    & Physical demand           & 1.50                  & 0.76                    & 1.25                  & 0.46                    & 0.157                                  \\
                                    & Temporal demand           & 1.88                  & 0.99                    & 1.38                  & 0.52                    & \textbf{0.046}                         \\
                                    & Effort                    & 2.88                  & 1.25                    & 3.13                  & 0.99                    & 0.680                                  \\
                                    & Performance               & 5.00                  & 1.07                    & 4.75                  & 0.89                    & 0.458                                  \\
                                    & Frustration               & 2.13                  & 1.46                    & 2.75                  & 1.28                    & 0.236                                  \\ \hline
\end{tabular}
\caption{Survey results comparing the baseline and \sysname{} conditions across multiple measures related to exploration and system interaction, with statistically significant differences (\textit{p} < 0.05) highlighted in bold. The measures includes responses on the accuracy of expressing exploration intent, system understanding, ease of adjustment, satisfaction with outcomes, and fixation. Aggregated Creative Support Index (CSI) and NASA-TLX workload measures are also presented.}
\end{table*}

\section{Examples of Generated Images from Two Conditions} \label{appendix:example}
\begin{figure*}
    \centering
    \includegraphics[width=\textwidth]{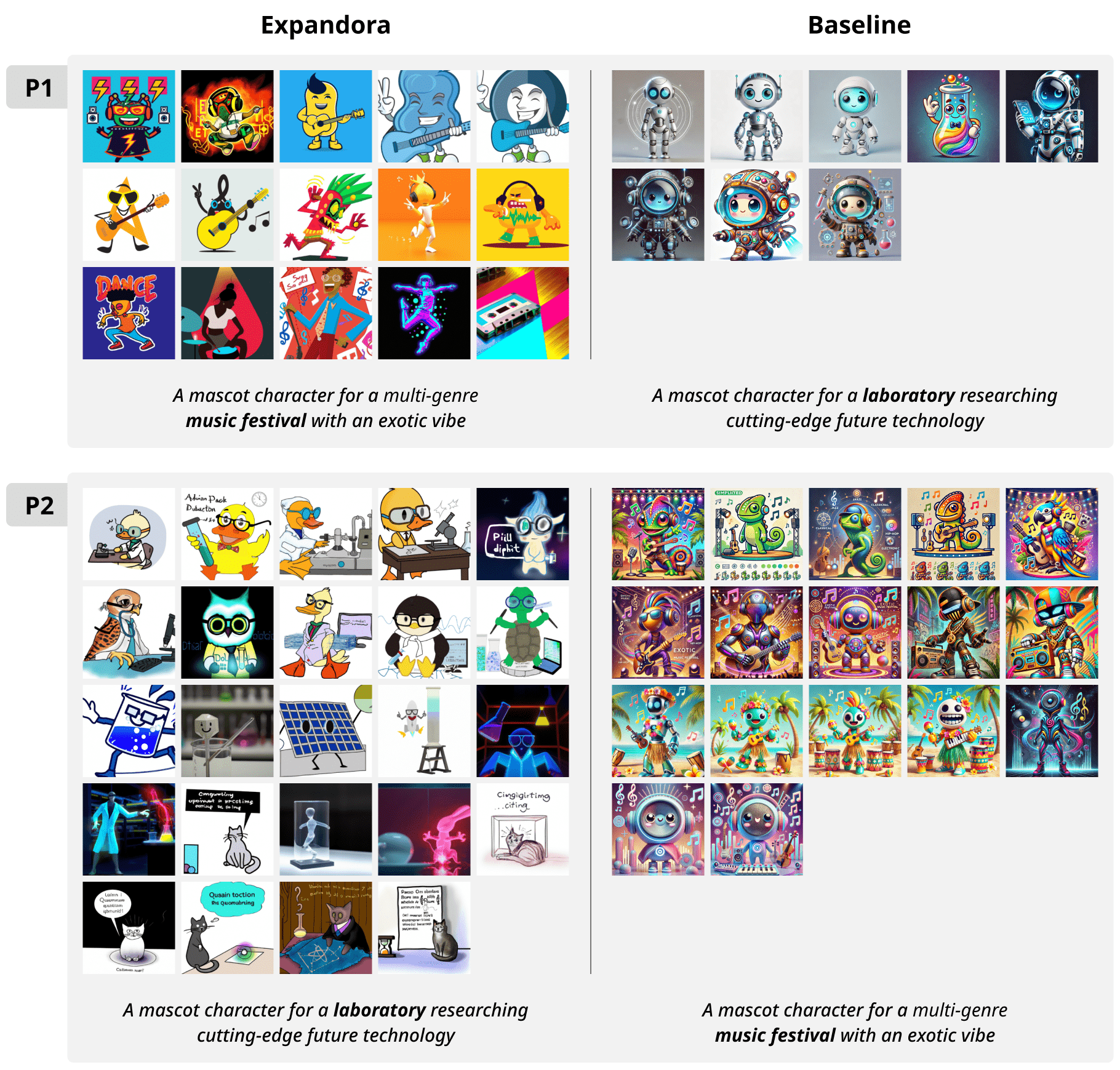}
    \caption{Examples of images generated by P1 and P2 using \sysname{} and the baseline. Images are displayed in the order they were generated, with the corresponding design topics noted below each set.}
    \label{fig:usage-example}
\end{figure*}

Figure~\ref{fig:usage-example} shows examples of images generated by two participants using \sysname{} and the baseline. In \sysname{} condition, users produced a wider variety of creative concepts, demonstrating the system's ability to support diverse and iterative exploration.

\end{document}